# ACDC: Altering Control Dependence Chains for Automated Patch Generation

Rawad Abou Assi　　　Chadi Trad　　　Wes Masri

Department of Electrical and Computer Engineering
American University of Beirut
Beirut, Lebanon
{ria21, cht02, wm13}@aub.edu.lb

## ABSTRACT

Once a failure is observed, the primary concern of the developer is to identify what caused it in order to repair the code that induced the incorrect behavior. Until a permanent repair is afforded, code repair patches are invaluable. The aim of this work is to devise an automated patch generation technique that proceeds as follows: *Step1*) It identifies a set of failure-causing control dependence chains that are minimal in terms of number and length. *Step2*) It identifies a set of predicates within the chains along with associated execution instances, such that negating the predicates at the given instances would exhibit correct behavior. *Step3*) For each candidate predicate, it creates a classifier that dictates when the predicate should be negated to yield correct program behavior. *Step4*) Prior to each candidate predicate, the faulty program is injected with a call to its corresponding classifier passing it the program state and getting a return value predictively indicating whether to negate the predicate or not. The role of the classifiers is to ensure that: 1) the predicates are not negated during passing runs; and 2) the predicates are negated at the appropriate instances within failing runs.

We implemented our patch generation approach for the Java platform and evaluated our toolset using 148 defects from the *Introclass* and *Siemens* benchmarks. The toolset identified 56 full patches and another 46 partial patches, and the classification accuracy averaged 84%.

## 1 INTRODUCTION

During the debugging process, the developer replicates the failure at hand in order to: 1) identify what caused it, and 2) prevent it from happening again by modifying, adding, or deleting code. These two activities are respectively termed *fault localization* and *program repair*. In most cases, these activities cannot be completed in a timely manner which calls for the temporary reliance on *automated patch generation*, the subject of this work.

For over three decades, researchers have proposed a plethora of automated fault localization techniques and tools, and in recent years a number of automated program repair and patch generation techniques have been proposed that leverage varying approaches such as evolutionary algorithms [19][39], constraint solving [14][16][11][26][27], and program mutation [10]. The aim of this work is to devise an effective patch generation technique that leverages a variant of an accurate coverage-based fault localization (CBFL) approach that was previously presented in the literature [1].

CBFL techniques generally entail two main steps. First, they identify the executing program elements that correlate most with failure. Second, starting from these elements, which are not necessarily the causes of the failure, they try to locate the faulty code following some examination strategy. It often happens that in the first step the correlation measure of the identified elements is not high enough to successfully guide the developer to the fault. This shortcoming is likely due to the fact that the program elements covered are simple, such as statements and branches, and therefore, cannot characterize most defects that are typically complex. This calls for covering program elements whose complexity matches the complexity of the defect under consideration, no more nor less. A less complex element cannot characterize the defect to begin with; whereas, an excessively complex element is likely to 'subsume' the defect and to successfully characterize it; but might lead to erroneously tagging too many statements as suspicious. The ultimate goal then is to define a program element type that characterizes as closely as possible the defect at hand. The CBFL technique presented in [1] attempts to achieve that goal by identifying the data/control dependence chains that correlate with failure, and are minimal in number and length.

Our patch generation approach uses a variant of the above CBFL work as a starting point. But it first improves its accuracy by considering the causal relationships amongst program statements. The proposed patch generation approach proceeds as follows:

***Step1***. It identifies a set of suspicious control chains via the improved CBFL technique.

***Step2***. It identifies a set of predicates within the suspicious chains along with associated execution instances such that negating the predicates at the given instances would exhibit correct program behavior.

***Step3***. A classifier is created for each candidate predicate whose purpose is to dictate when the predicate should be negated to yield correct behavior. The training output data for the classifier (to negate vs. not to negate) is deduced from the execution instances identified in *Step2*. The training input data is derived from the program state captured at the point of predicate execution.

***Step4***. Prior to each candidate predicate, the faulty program is injected with a call to its corresponding classifier. At runtime, the call passes the classifier the program state and returns the predicted decision on whether to negate the predicate or not.

In summary, the proposed approach aims at generating software patches as a result of which failing test cases would pass. Developers will likely opt to temporarily use the patches until a permanent manual repair is afforded. Furthermore, while seeking manual repairs, the minimal sets of predicates to be negated can be invaluable for the developers.

The main contributions of this work are:

a. An effective patch generation toolset comprising two components:
   i. A highly accurate CBFL component that identifies failure-causing control dependence chains.
   ii. A patch generation approach centered on altering the identified control chains.
b. An implementation of the toolset that targets the Java platform.
c. An evaluation of the toolset demonstrating its effectiveness at patching 148 faults.

Section 2 describes the approach for identifying control dependence chains that are failure causing. Section 3 describes how the chains are altered for the purpose of patch generation. In Section 4 our patch generation approach is evaluated by applying it onto 12 subject programs involving 148 defects. Section 5 discusses threats to the validity of our approach. Section 6 surveys related work, and Section 7 concludes.

## 2  *CBFL*: IDENTIFYING FAILURE-CAUSING CONTROL DEPENDENCE CHAINS

Numerous fault localization techniques based on coverage of simple program elements such as statements and predicates have been presented in the literature [3][17][34][20][21]. In the case of a defect that involves complex interactions between many program elements, e.g., a combination of def-use pairs and branches executed in some specific order, these techniques are not likely to locate the faulty statements [12]. Several researchers tried to address this issue by using more complex program elements [36][24][1], and the CBFL work presented in [1] uses program elements that vary in complexity in order to better match the complexity of the defect at hand, namely, dependence chains with varying lengths. The first component in our patch generation toolset is a highly accurate CBFL technique that is an extension of the work presented in [1]. Next we describe our CBFL component while focusing on the extended parts of the work.

### 2.1  Definitions

This section provides background definitions that supersede the ones presented in [1].

**Definition** – A ***direct control dependence*** is a couple $(s_1, s_2)$ that satisfies the following: 1) $s_1$ is a predicate statement; 2) $s_2$ is a statement that is directly control-dependent on $s_1$.

**Definition** – A ***direct control dependence*** $(s_1, s_2)$ is said to be ***executed*** by a test case $t$ if $s_2$ is executed by $t$.

**Definition** – A ***chain*** is a sequence of nodes $(s_1, s_2, ..., s_k)$ where $k \geq 2$ such that $(s_i, s_{i+1})$ is a direct control dependence $\forall\ 1 \leq i \leq k\text{-}1$.

**Definition** – A ***chain*** $(s_1, s_2, ..., s_k)$ is said to be ***executed*** by a test case $t$ if there exists a sequence of time instants $tm_1, tm_2, ..., tm_{k-1}$ such that:

1. $0 < tm_1 < \cdots < tm_{k-1}$
2. $\forall\ 1 \leq i \leq k\text{-}1$, the direct control dependence $(s_i, s_{i+1})$ is executed by $t$ at $tm_i$.

**Definition** – Given a chain $c = (s_1, s_2, ..., s_k)$, we denote by ***head(c)*** the statement $s_1$ and by ***tail(c)*** the statement $s_k$. We define ***length(c)*** to be $k\text{-}1$. $\forall\ 1 \leq i \leq k\text{-}1$, $s_i$ is said to be the predecessor of $s_{i+1}$.

**Definition** – A chain $e = (s_1, s_2, ..., s_k)$ is said to be an ***extension*** of another chain $c = (r_1, r_2, ..., r_p)$ iff $k > p$ and $\forall\ 1 \leq i \leq p$, $r_i = s_i$.

**Definition** – A chain $c$ is said to be ***maximal*** in a set of chains $S$ iff no extension of $c$ is contained in $S$.

### 2.2  Identifying Failure-Correlated Control Dependence Chains

We now describe the basic high level steps for identifying failure-correlated control dependence chains, which mirror the steps presented in [1]:

1. **Specify *C* to represent control chains of length one**.
2. **Compute the suspiciousness metric *M* for all executing chains.** This step involves executing a test suite on the subject program in order to collect execution profiles describing the frequency of occurrences of each chain, which calls for invoking *Algorithm-2* provided in [1]. This step is in essence similar to what is done in [17], except that it differs in the type of the program elements considered. Note that *M* represent the *Ochiai* metric presented in [2].
3. **Exit if any number of chains exhibited a score of 1.0.** The algorithm terminates if one or more chains exhibited high correlation with failure. Note that chains not sharing any statements between failing and passing runs are discarded. Such measure aims at reducing the rate of false positives by focusing on the failing and passing profiles that are most similar [34].
4. **Increase the complexity of *C*** by increasing the length of the control chain by one.

5. **Exit if the complexity of *C* renders profile collection infeasible, otherwise go to step *2***. Profile collection is considered infeasible when its duration exceeds a certain threshold, for a given chain length.

These steps differ from their counter-parts presented in [1] in the following: 1) they do not explore program statements; 2) they only explore control chains as opposed to control/data chains under the assumption that the former enables simpler repairs; and 3) the resulting suspicious chains are refined by considering the causal relationships amongst program statements, as described in Section 2.3.

The premise is that the above algorithm will identify a small set of control chains that are highly correlated with failure, i.e., having a high *M* value, preferably equal to 1.0. This approach is likely to be more effective than other techniques that are based on covering a non-varying type of program elements. If the elements covered by such techniques were simpler than the defect, it is likely that no highly failure-correlated elements would be found, whereas, if the elements were more complex than the defect, it is likely that too many would be found.

## 2.3 Pinpointing Failure-Causing Control Dependence Chains

In this section we first provide background on *causal inference* [31][32][35] then describe how the chains identified in Section 2.2 are further refined.

### 2.3.1 *Causal Inference: Background*

A common misconception is that if two variables are correlated, then one causes the other, i.e., a cause-and-effect relationship connects them. In fact, correlation does not imply causation, but causation implies non-linear correlation. Causality is clearly more desirable than correlation for the purpose of fault localization, since the ultimate goal is to identify and repair the code that caused the failure and not just any code that correlated with it. Early CBFL technique (erroneously) used correlation to compute the suspiciousness score in order to infer the causal effect of individual program elements on the occurrence of failure. The scores they used suffer from *confounding bias*, which occurs when an *apparent* causal effect of an event on a failure may actually be due to an unknown confounding variable, which causes both the event and the failure. Confounding bias might explain the high rate of *false positives* exhibited by such techniques [29].

Given a program and a test suite, assume for example that all failing test cases induce dependence chain $e_1 \rightarrow e_2 \rightarrow e_{bug} \rightarrow e_3 \rightarrow e_4 \rightarrow e_{fail}$ and all passing test cases induce $e_1 \rightarrow e_2$ only; where $e_{bug}$ exercises the fault and $e_{fail}$ indicates a failure. A correlation-based approach would determine that any of $e_{bug}$, $e_3$, or $e_4$ is equally suspect to be the cause of the failure, thus resulting in two false positives. Whereas, a causation-based approach that considers dependences to have causal effect, would determine that $e_4$ is the least suspect and $e_{bug}$ the most suspect. This is because: 1) confounding bias weakens the causal relationship; and 2) when computing the suspiciousness scores, the confounding bias to consider for $e_4$ would involve $e_3$ and $e_{bug}$, for $e_3$ it would involve $e_{bug}$, and no confounding is involved when computing the suspiciousness score of $e_{bug}$.

Confounding bias is a common phenomenon that needs to be identified, controlled, and reduced. Pearl's *back-door criterion* [31][32] allows for graphically identifying confounding bias and for reducing it by employing causal effect estimators [35]. In case we wanted to measure the causal effect of program element *pe* on the program outcome *Y*, we need to perform the following steps: 1) build a causal graph around *pe* and *Y*, possibly derived from prior analyses such as dependence analysis; 2) use the *back-door criterion* method to identify the *back-door paths* in the graph and their corresponding covariates to control; and 3) devise an estimator for the suspiciousness of *pe* using coverage information about *pe*, the controlled covariates, and the program outcome. Assume that the first step yielded a causal graph with the three edges {*pe*→*Y*, *pe'*→*pe*, *pe'*→*Y*}, the *back-door criterion* would identify *pe*←*pe'*→*Y* as the single *back-door path* in the graph, and therefore would deem *pe'* as the covariate to control (since it causes both *pe* and *Y*).

In summary, the intuition behind the back-door criterion is as follows. The back-door paths in a causal graph carry spurious associations from *pe* to *Y*, while the paths directed along the arrows from *pe* to *Y* carry causative associations. The goal is to block the back-door paths in order to ensure that the measured association between *pe* and *Y* is purely causative.

### 2.3.2 *Causal Inference for CBFL*

Baah, Podgurski, and Harrold were the first to investigate the application of causal *inference* in CBFL [5]. Given a statement *s* in program *P*, the aim of their work is to obtain a causal-effect estimate of *s* on the outcome of *P* that is not subject to severe confounding bias, i.e., a causation-based suspiciousness score of *s*. They applied Pearl's Back-Door Criterion to program control dependence graphs in order to devise an estimator based on the following linear regression model:

$$Y_s = \alpha_s + \tau_s T_s + \beta_s C_s + \varepsilon_s$$

This model relates the event of program failure $Y_s$ with not just the event of covering statement *s* (i.e., $T_s$), but also with the confounding events, listed in $C_s$. The *model* is fit separately for each statement *s*, using statement execution profiles that are labelled as passing or failing.

Given that causal graphs are not known in practice, this work assumes that: 1) if *s* is faulty, covering it will cause a failure; and 2) if *s* is dynamically directly control dependent on statement *s'*, *s'* causes the execution of *s* and possibly the failure, i.e., *s'* is the only source of confounding bias and $C_s$ becomes a single indicator of whether *s'* was covered. In other words, this work assumes that the *causal* graph is made up of the following three edges only: $T_s \rightarrow Y_s$, $C_s \rightarrow T_s$, and $C_s \rightarrow Y_s$. Therefore, the only back-door path to be controlled is $T_s \leftarrow C_s \rightarrow Y_s$. More importantly, since $\tau_s$ is the average effect of $T_s$ on $Y_s$, the work uses an estimate of $\tau_s$ to quantify the failure-causing effect of *s*, i.e., the suspiciousness of *s*.

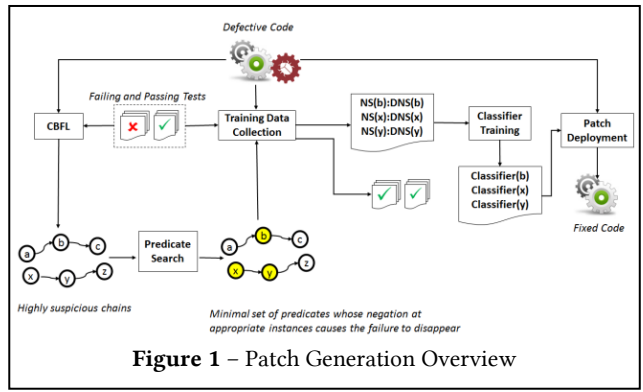

**Figure 1** – Patch Generation Overview

Note that neither of the above assumptions is sure to hold. For example, due to coincidental correctness, covering a faulty statement will not necessarily cause a failure; and due to the transitivity of control/data dependences in programs, direct control dependences might not be the only source of confounding bias. Nevertheless, using the above model is likely to yield more accurate suspiciousness scores than simply relying on correlation. In fact, the empirical study conducted in [5], which involved 9 subject programs and 168 faulty versions, indicated that the proposed causal-effect estimator can significantly improve the effectiveness of fault localization over existing correlation-based metrics.

Finally, Baah et al. [6] later improved the above approach by considering patterns of dependences; [37] proposed the use of other causal inference techniques in method-level fault localization; and [7] developed a fault localization technique focused on numerical software using a value-based causal model.

### 2.3.3  *Refining the Chains for Improved Repair*

Refining the chains identified in Section 2.2 is an important step to arrive at a more accurate fault localization which eventually leads to a more effective patch generation. To do so, we first compute the causal effect for each statement *s* appearing in the chains previously identified using the approach adopted in [5] as follows:

1. Fit a linear regression model in the form of $Y_s = \alpha_s + \tau_s T_s + \beta_s C_s + \varepsilon_s$ as discussed in Section 2.3.2. In our case, $C_s$ represents the statement upon which *s* is directly control-dependent. That is, $C_s$ would be 1 if a given test cases covers the control predecessor of *s* and 0 otherwise. If *s* has no control predecessor, the model to fit would be $Y_s = \alpha_s + \tau_s T_s + \varepsilon_s$.
2. The causal effect of *s* is estimated via the coefficient $\tau_s$.

To refine the chains resulting from the steps presented in Section 2.2, we sort them based on the maximum causal effect per chain and select the top three.

## 3  ALTERING THE CONTROL CHAINS

Our patch generation approach is based on the premise that a defect is likely to trigger erroneous branch executions within highly suspicious control chains. As such, we expect that properly altering such chains at runtime is likely to cause the failure to disappear. This section describes how we alter the chains obtained from the CBFL phase described in Section 2 to arrive at a patch for the faulty program.

### 3.1  Overview

Figure 1 provides an overview of our approach. CBFL is first applied to obtain the suspicious control chains. The ***Predicate Search*** algorithms are invoked to identify a minimal set of predicates whose negation at proper instances causes the failure to disappear in all test cases that were originally failing. In the ***Training Data Collection*** phase, the test suite is executed in order to capture the program states relevant to each candidate predicate *p*, every time *p* is executed. The captured states are grouped into two sets: those associated with when *p* needs to be negated (*NegateState(p)* or *NS(p)*) and those obtained when a negation is not required (*DontNegateState(p)* or *DNS(p)*). ***Classifier Training*** involves using *NS* and *DNS* to train a non-linear SVM classifier that decides whether or not to negate *p*. In the ***Patch Deployment*** phase, the resulting classifiers, one per candidate predicate, are integrated within the defective source code to arrive at a patched version of the program. Next we describe the *Predicate Search* phase and the *Training Data Collection* phase; the other two phases are not as valuable in terms of their contribution.

### 3.2  Predicate Search

We devised two complementary search algorithms that identify which predicates to negate and when to negate them. We recognize five scenarios that our algorithms will likely encounter:

> ***Scenario1.*** A candidate predicate that is exercised by only failing runs should be negated at all times. This is clearly a simple case in which our approach can tackle without the need of a classifier.
> ***Scenario2.*** A candidate predicate that is exercised by both passing and failing runs should be negated all the time only during failing runs. For that, our proposal to use a classifier is critical as it will insure that the predicate will be negated during failing runs only.
> ***Scenario3.*** A candidate predicate that is exercised by both passing and failing runs should be negated at specific occurrences (but never all the time). Since such occurrences and their number might vary from one failing run to another, we need to identify a *pattern* that abstracts when the negation must occur across all failing runs. Consequently, the classifier must insure that the predicate will be negated according to that pattern of execution.
> ***Scenario4.*** A candidate predicate that is exercised by both passing and failing runs should be negated at specific occurrences or all the time.

```
SinglePredicateSearch(PredList_susp, T_fail)

1. Patterns = {"all", "first", "last", "all-first", "all-last",
    "all-(first+last)", "first+1", "last-1", "first+last", "odd", "even"}
2. PredList_solution = ∅

3. for each p in PredList_susp  do
4.   for each t_fail in T_fail do
5.     for each pattern in Patterns do
6.       execute t_fail while negating p according to pattern
7.       if execution succeeds
8.         p.repairs(t_fail, pattern)
9.         PredList_solution = PredList_solution ∪ p
         endif
       endfor
     endfor
endfor

// find a predicate that repairs all failing tests following a single pattern
10. for each p in PredList_solution do
11.   for each pattern in Patterns do
12.     if |p.getRepairedTestsByPattern(pattern)| = |T_fail|
13.       return (p, pattern)
         endif
     endfor
   endfor
endfor

return null
```

**Figure 2** – Single Predicate Search Algorithm

*Scenario5.* The full set of failing test cases can exhibit correct behavior only if multiple predicates are negated, such that each predicate follows a single pattern. That is, patching a given defect involves multiple predicates and classifiers, at different locations in the code.

The first three scenarios are supported by the *SinglePredicateSearch* algorithm shown in Figure 2, and the fourth scenario is supported by the *MultiplePredicateSearch* algorithm shown in Figure 3.

*SinglePredicateSearch* considers a single suspicious predicate $p$ at a time and eleven different patterns of negation. Specifically, it checks whether any of the following actions would make all the failing test cases succeed: 1) negating $p$ all the time within a given failing test case; 2) negating $p$ the first time; 3) negating $p$ the last time; 4) negating $p$ all the time except the first; 5) negating $p$ all the time except the last; and so on, as indicated on Line 1 of the pseudocode.

*SinglePredicateSearch* takes as input: 1) $PredList_{susp}$: the list of suspicious predicates identified by the CBFL component; and 2) $T_{fail}$: the set of failing test cases within the training set. Line 1 initializes *Patterns* with the execution patterns to be matched; note that the patterns are roughly ordered in terms of their simplicity. On Line 2, **PredList_solution** is initialized to the empty set; its role is to store the suspicious predicates that are candidates for repairing one or more failing runs. For every suspicious predicate $p$, every failing test $t_{fail}$, and every pattern *pattern*, Line 6 executes $t_{fail}$ while negating $p$ according to *pattern*. In case the execution succeeds, $p$ is deemed to be a viable candidate for repairing $t_{fail}$ according to *pattern*. Accordingly, Line 8 associates $p$ with $t_{fail}$ and *pattern*, and Line 9 adds $p$ to **PredList_solution**. Lines 10-13 search **PredList_solution** for a predicate that repairs all tests in $T_{fail}$ according to the same pattern, and returns the first one found. In case *SinglePredicateSearch* failed to find and return a solution at Line 13, *MultiplePredicateSearch* is invoked. (Note that in Section 4.2 we use a slightly modified version of *SinglePredicateSearch* that returns all possible solutions, as opposed to just the first encountered one.)

*MultiplePredicateSearch* is a greedy algorithm that searches for a set of predicates that collectively repair all failing test cases such that each predicate follows a single pattern; Figure 3 presents the corresponding pseudocode. Lines 1-5 initialize a 2-dimensional array where rows represent failing tests and columns represent suspicious predicates. The value at location $(i, j)$ would be *true* if the $i^{th}$ test case is fixed by negating the $j^{th}$ predicate and *false* otherwise. The loop starting at line 8 repeatedly selects a predicate $p_{max}$ to be included in the solution as follows. Line 9 identifies $p_{max}$ as the predicate that fixes the largest number of failing tests among those that could not be fixed using the predicates chosen in previous iterations. Lines 10-11 add $p_{max}$ to the result and update the number of failing tests that are fixed by the current solution. Lines 12-15 update the matrix such that all the failing tests that are fixed by $p_{max}$ would not be considered in subsequent iterations.

## 3.3 Training Data Collection

Given a pair ($p$, *pattern*) identified by *SinglePredicateSearch* or *MultiplePredicateSearch*, we now need to collect data to train a classifier that will guide the execution by indicating when to negate $p$ according to *pattern*. Two sets of data are actually needed, *NegateState(p)* which is associated with when $p$ needs to be negated and *DontNegateState(p)* which is associated with when $p$ should remain intact. The two sets are built by collecting the approximated state of the program right before each execution of $p$. Specifically, on the onset of $p$ executing, the values of the following program variables are collected:

1) *Use*($p$), i.e., the variables or method return values directly *used* in $p$.
2) Formal parameters of the method containing $p$.
3) Local and static variables that were *used* or *defined* within the method containing $p$.
4) Object attributes that were *used* or *defined* within the method containing $p$.

The values of the program variables are derived according to their types, as follows:

1) Variables of scalar types (e.g., *int*, *float*, *char*) have their values used as is.
2) *java.lang.String* objects are represented by their *hashCode*() values.
3) The values of the non-String objects are computed by considering the states of their attributes, and if need be, by recursively considering the attributes of their attributes and so on. In other words, to represent an object, a scalar was derived based on all of its direct and indirect attributes (similar to how an accurate *hashcode* would be computed).

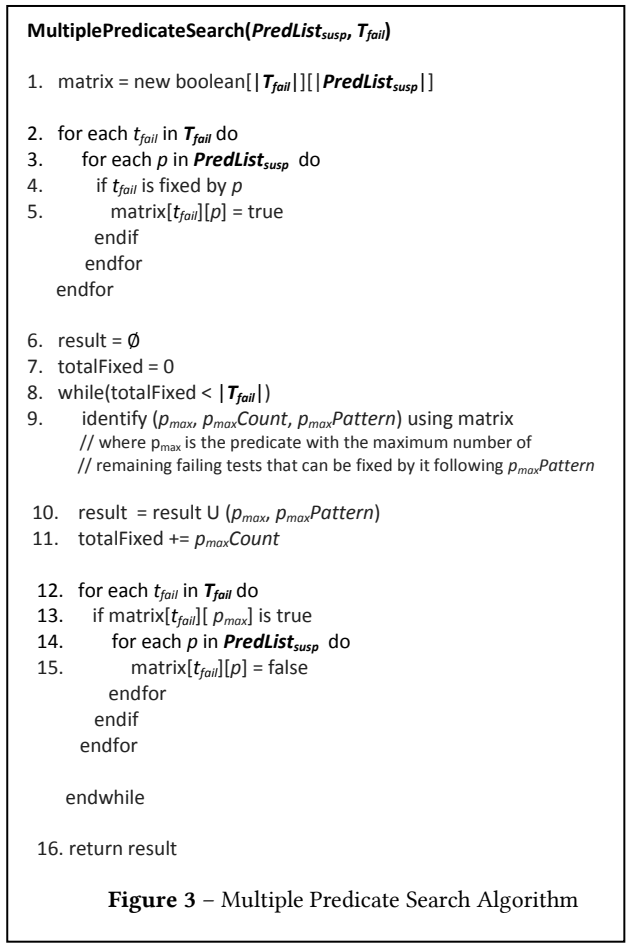

**Figure 3** – Multiple Predicate Search Algorithm

## 3.6 Implementation

Our implementation targeted the Java platform at the byte code level. Part of the work that posed most implementation challenges included *Predicate* Search, and *Training Data Collection* which both involved instrumenting and profiling Java byte code using the *Byte Code Engineering Library*, BCEL.

*Training Data Collection* calls for developing a *state* profiling engine that captures a snapshot of the approximated program state at given code locations, as described in Section 3.3 The profiling engine consists of two main subcomponents: the *Instrumenter* and the *Profiler*. The preliminary step is to instrument the target byte code class files using the *Instrumenter* which inserts a number of method calls to the *Profiler* at given points of interest. At runtime, the instrumented application invokes the *Profiler*, passing it information that enables it to log the approximated program states.

*Predicate Search* aims at identifying which predicates to negate and following which pattern. In order to enforce a given pattern, the number of occurrences of each suspicious predicate needs to be known a priori, thus requiring each failing test to be executed twice. To support the conditional negation of a given predicate (Line 6 of *SinglePredicateSearch*), each suspicious predicate is augmented by byte code that enables the *Predicate Search* profiler to negate the predicate when needed.

Figure 4 is an illustration of how a predicate would be instrumented. The ID of the predicate is pushed on the stack (Line 8) and a method shouldNegate() is invoked (Lines 9-10) right before the if statement executes. The returned value is checked (Line 11). If the returned value is *false*, the code branches to the original if statement (Line 14). If the returned value is *true*, a synthesized if statement with swapped branches is executed (Lines 12-13). On the profiler side, shouldNegate() logs the timestamp and the ID of the executed if statement, and returns *true* or *false* according to the algorithms described in Section 3.2.

## 4 EMPIRICAL EVALUATION

The first research question we aim to answer is: (RQ1) "*How effective are our Predicate Search algorithms at finding candidate predicates?*" Recall that these predicates provide the basis for patch generation. The second question we want to answer is: (RQ2) "*How effective is state profiling at building accurate classifiers?*" Note that other options include profiling of structural program elements, such as *branches*, *def-uses*, or even dependence chains. In order to address these questions we applied our toolset to 12 Java programs for a total of 148 defects. Next we describe the used subject programs then present and discuss our results.

### 4.1 Subject Programs

Our experiments involved 57 defective versions from the *Siemens* benchmark (*sir.unl.edu*) and 91 versions from the *Introclass* benchmark [19]. The Siemens versions, namely, 8 *print_tokens2* versions, 4 *print_tokens* versions, 6 *replace* versions, 4 *schedule* versions, 18 *tcas* versions, and 17 *tot_info* versions were manually converted to Java in [1]. The *Introclass* benchmark is originally written in C. It contains 7 programs (*digits*, *grade*, *median*, *smallest*, *syllables*, and *checksum*) and hundreds of related bugs. We opted to randomly select 20 versions from each program and convert them to Java. However, we excluded redundant bugs, those whose Java versions did not fail, and those whose Java versions caused exceptions to be thrown. As a result, we used 20 *digits* versions, 20 *grade* versions, 20 *median* versions, 20 *smallest* versions, 4 *syllables* versions, and 7 *checksum* versions for a total of 91.

| // Original Code | // Instrumented Code |
|---|---|
| if (x<0) // identified as 1001<br>    x=x+1;<br>else<br>    y=y+1;<br>z=z+1; | if ((x<0) ^ shouldNegate(1001))<br>    x=x+1;<br>else<br>    y=y+1;<br>z=z+1; |
| // Original Byte Code | // Instrumented Byte Code |
| 7  iload_1      14 iload_2<br>8  ifge 14     15 iconst_1<br>9  iload_1      16 iadd<br>10 iconst_1    17 istore_2<br>11 iadd         18 iload_3<br>12 istore_1    …<br>13 goto 18 | 7  iload_1         17 iadd<br>**8**  **sipush 1001**   18 istore_1<br>**9**  **invokestatic**    19 goto 24<br>**10** **shouldNegate(I)Z** 20 iload_2<br>**11** **ifeq 14**        21 iconst_1<br>**12** **ifge 15**        22 iadd<br>**13** **goto 20**        23 istore_2<br>**14** **ifge 20**        24 iload_3<br>15 iload_1        …<br>16 iconst_1 |

**Figure 4** – Instrumentation for Negating Predicates

## 4.2 RQ1: *How effective are our Predicate Search algorithms at finding candidate predicates?*

We applied the CBFL component on the 148 defective versions of our subject programs in order to produce for each a corresponding list of suspicious predicates ($PredList_{susp}$). We then applied the *SinglePredicateSearch* algorithm on each version to identify a single (predicate, pattern) pair that could be used as the basis for a patch. In case no solution was found by *SinglePredicateSearch*, we applied the *MultiplePredicateSearch* algorithm which aims at finding a set of predicates that collectively could be used as the basis for a patch. To better analyze the results of our experiments, we slightly modified *SinglePredicateSearch* so that it produces not just the first encountered solution (see Line 13 in Figure 2) but all possible solutions. We also modified it to produce partial solutions, i.e., pairs based on which not all (but some) failing test runs could be patched.

The outcome of applying *SinglePredicateSearch* and *MultiplePredicateSearch* on the *Introclass* versions is summarized below:

1) Using *SinglePredicateSearch*, at least one solution was found for 20 out of the 91 versions. Actually, for most of the 20 versions, there were more than one solution. For example, *digits_v11* had 6 solutions involving two different predicates and four different patterns: ($p_1$, "all"), ($p_1$, "first"), ($p_1$, "first+last"), ($p_1$, "odd"), ($p_2$, "first"), ($p_2$, "odd").

2) Using *MultiplePredicateSearch*, a solution was found for an additional 8 versions. For example, *digits_v1* had a solution involving two predicates: {($p_1$, "all"), ($p_2$, "first")}, and *grade_v5* had a solution involving five predicates: {($p_1$, "all"), ($p_2$, "all"), ($p_3$, "all"), ($p_4$, "all"), ($p_5$, "all")}.

3) In addition to the above 28 full patches, *SinglePredicateSearch* identified 37 partial patches; i.e., patches that fix some of the failing test cases but not all of them. For example, 4 out of 12 failing tests were fixed for *grade_v4* using one predicate paired with either one of the following patterns: "all", "first", "first+last", "odd".

Applying the *Predicate Search* algorithms on the Siemens versions yielded the following results:

1) At least one solution was found for 21 out of the 57 versions using the *SinglePredicateSearch* algorithm. For example, *tcas_v27* had 11 solutions involving three predicates: ($p_1$, "first"), ($p_1$, "first+last"), ($p_1$, "odd"), ($p_2$, "all"), ($p_2$, "first") ($p_2$, "first+last"), ($p_2$, "odd"), ($p_3$, "all"), ($p_3$, "first") ($p_3$, "firstlast"), ($p_3$, "odd").

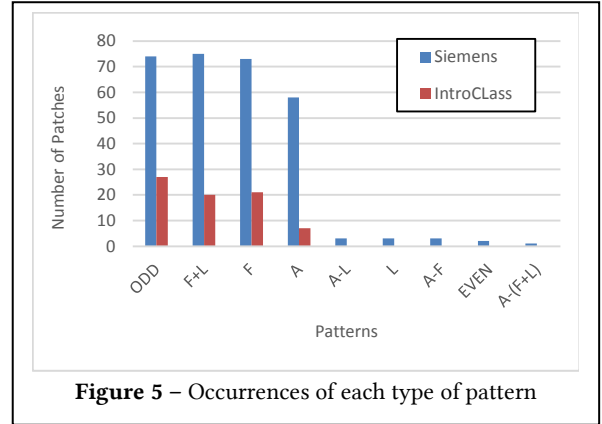

**Figure 5** – Occurrences of each type of pattern

2) A solution was found for an additional 7 versions using *MultiplePredicateSearch*. For example, *tcas_v2* had a solution involving two predicates: {($p_1$, "all"), ($p_2$, "all")}.

3) Furthermore, *SinglePredicateSearch* identified 9 partial patches in addition to the 28 full patches. For example, 9 out of 50 failing tests were fixed for *print_tokens2_v1* using one predicate paired with either "even" or "all-first".

Further observations can be made about our *Predicate Search* results:

1) Given the versions in *Introclass* having at least one full patch, the min, max, and average number of patches per version are 1, 8, and 4.75, respectively. The corresponding metrics for the *Siemens* versions are 1, 44, and 14.31.

2) Figure 5 shows the number of times each of the eleven patterns was involved in a patch identified by *SinglePredicateSearch* (in case a version has multiple solutions, all solutions are considered). Clearly, "all", "first", "odd", and "first+last" are the predominant patterns.

3) Figure 6 shows the distribution of the identified patches across the five scenarios we listed in Section 3.2. A typical naive repair approach would count on the fact that *Scenario1* would be the most prevalent. However, it is apparent from the figure that our patches mostly involved the other scenarios that are more complex. This observation indicates that our approach is more effective.

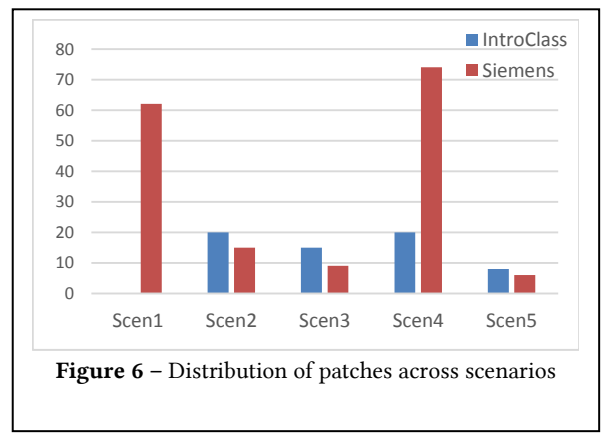

**Figure 6** – Distribution of patches across scenarios

## 4.3 RQ2: *How effective is state profiling at building accurate classifiers?*

Long and Rinard [22] distinguish between two types of patches: a) "plausible" patches that pass all the test cases in the validation suite and b) "correct" patches that generalize to test cases outside the validation suite. To evaluate the generalizability of our classifiers, we conducted the following experiment for each of the patches identified by our approach:

1) Randomly partition the test suite into 2 sets (*training* and *testing*).
2) Build a non-linear SVM using the states captured within the *training* group at the execution point(s) of the candidate predicate(s).
3) Use the classifier from step 2) to fix the failing tests in the *testing* group.
4) Run the passing tests of the *testing* group under the classifier to check whether the latter has a negative impact on the correct behavior of the program.
5) Measure the accuracy of the classifier as the ratio of test cases that were handled properly among the *testing* group, i.e. failing tests that got fixed and passing tests that remained intact.

Figure 7 shows a boxplot of the measurements we obtained while experimenting with training set sizes of 5%, 10%, 20%, 40%, and 80%. The average classification accuracy for the five training sets were respectively 79.4%, 79.3%, 86.1%, 87.9%, and 86.8%. Using a Wilcox test with $p<0.05$, we found that the difference between the training sets is not statistically significant.

In addition, we considered a hypothetical scenario in which we used the entire test suite for training and then used it for testing as an indicator of the maximum accuracy that can be achieved using our current implementation. The accuracy ranged between 23% and 100% with an average of 90.4%. This leads us to the conclusion that state information (as we approximated it) is in most cases (but not all) sufficient to characterize the conditions of predicate negation.

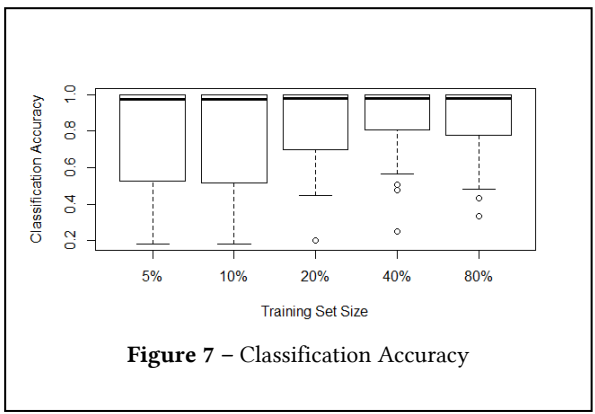

**Figure 7** – Classification Accuracy

## 4.4 Discussion

Given that our *Predicate Search* algorithms yielded 56 full patches and 46 partial patches for the 148 faulty versions. One would consider our approach to be effective to a large extent, i.e., the answer to RQ1 would be "reasonably effective". Meanwhile, we believe that there is ample room for improving our search approach by additionally targeting bugs that cannot be fixed via altering control flow, and by exploring more sophisticated search techniques, the subject of future work.

Our state profiler operates at a high level of granularity; therefore we expected it to yield highly accurate classifiers. The results in Section 4.3 support our expectation but not to the greatest extent, i.e., the answer to RQ2 would also be "reasonably effective". For example, a training set sized at only 5% that yields 79.4% accuracy, is impressive; but we expected higher accuracy measures for the larger training sets. Devising a new profiling technique that would improve the classification accuracy of our patch generation approach will be our top priority for the near future. Such technique will likely involve structural and temporal information in addition to state information.

## 5 THREATS TO VALIDITY

A major threat to the external validity of our approach is the fact that our experiments involved a limited number of subject programs and faults; therefore, it is not possible to draw broad conclusions based on our results. This could be remedied by conducting further experiments involving a variety of other subject programs from different domains and environments containing real and/or seeded defects.

We recognize the following threats to the internal validity of our approach:

1) Our CBFL approach assumes that most defects could be characterized by a few control dependence chains of some limited length. Actually, it is plausible that some defects might not be characterized by any structural profiling elements no matter how complex they are. Such defects are more likely to be characterized by profiling elements that capture the state information of the program.
2) Similar to the above concern, our approach assumes that most defects could be repaired by altering the control flow of a program. When such assumption does not hold, we should explore altering the program's data flow and devise state altering mutation operators.

As it is the case for most test suite based fault localization and program repair techniques, the effectiveness of our technique is greatly dependent on the quality of the test suite. This applies to both, identifying the suspicious chains and to training the classifiers. One way to tackle the issue of test suite quality is to leverage automated test case generation.

## 6 RELATED WORK

This section briefly discusses work related to coverage-based fault localization then discusses work related to patch generation and program repair.

## 6.1 Coverage Based Fault Localization

Jones et al. [17] presented a technique that uses visualization to assist with locating faults. They implemented their technique in a tool called Tarantula. The technique uses color and brightness to visually map the participation of each program statement in the outcome of the execution of the program with a test suite, consisting of both passing and failing test cases. To provide the visual mapping, the program statements are colored using a continuous spectrum from red to yellow to green: the greater the percentage of failing test cases that execute a statement, the brighter and more red the statement should appear.

Denmat et al. [12] studied the limitations of the technique presented by Jones et al. (2001). They argued that for it to be effective, the following three requirements must hold: 1) a defect is due to a single faulty statement, 2) statements are independent of each other, and 3) executing a faulty statement leads most of the time to a failure. Clearly, the aforementioned requirements are not likely to be fulfilled when dealing with complex programs involving non trivial defects.

Renieris and Reiss [34] described a technique that produces a report of the "suspicious" parts of a program by analyzing the spectra differences between the faulty run and the correct run that most resembles it. The experiments they conducted used basic block coverage spectra whereas the technique proposed here involves a much more complex spectra based on control dependence chains. Liblit et al. [20] and Liu et al. [21] each proposed a statistical fault localization method that uses coverage of some types of predicates.

Zhang et al. [43] described a technique that reduces the size of a dynamic slice based on analyzing the values taken by the variables involved in the slice execution. Xie and Notkin [41] analyzed value spectra differences to identify possible faults in modified programs. Agrawal et al. (1995) proposed a fault localization technique based on dicing; a dice being the set difference of two slices. Their technique is based on the assumption that the fault resides in the execution slice of a failing run but does not reside in the execution slice of a passing run. Dallmeier et al. [9] presented a tool for Java programs that locates likely failure-causing classes by comparing method call sequences of passing and failing runs. Clause and Orso [8] presented a technique for debugging failures that occur while the software runs on user platforms. Their technique allows for recording, replaying, and minimizing user executions. The resulting minimized execution can then be used to debug the defect(s) leading to the observed failure

## 6.2 Patch Generation and Program Repair

Zhang et al. [42] presented a fault localization technique that is very relevant to our patch generation approach. It entails switching the valuation of the program's predicates, each one at a time for the purpose of producing the correct behavior. A predicate switch that yields a successful program completion can be further analyzed in order to identify the cause of the defect. Our approach differs in that: 1) due to our accurate CBFL technique, only few predicates need to be explored for switching; 2) predicate switching is considered at execution instances discovered by our approach; and most importantly 3) a program patch supported by a classifier is provided.

Le Goues et al. [18] proposed *GenProg*, a repair technique based on genetic programming. They assume that repairing a fault in one function can make use of snippets of code appearing in other functions in the program. For example, several existing functions in a program might implement checks for whether a pointer is *null*, the corresponding code can then be inserted in the function under repair in the aim of repairing it. The technique explores different variations of the defective program such as those resulting from inserting statements, deleting statements, and swapping statements. Also, mutation and crossover operators are applied and guided using a fitness function that evaluates the generated program against the test suite. Once a repair is found, it is further refined using delta debugging by discarding the unnecessary statements within. Our repair technique is very different in terms of its underlying approach and the nature of the produced solution.

Recently, Assiri and Bieman [4] evaluated the impact of ten existing CBFL techniques on program repair. Specifically, they measured their impact on the effectiveness, performance, and repair correctness of a brute force program repair tool, i.e., a tool that exhaustively applies all possible changes to the program until a repair is found. A brute-force repair tool is guaranteed to fix a fault if a repair is feasible. Therefore, a failure to find a potential repair would likely be related to the selected CBFL technique. Including our proposed CBFL technique in their comparative evaluation would be valuable, as it could help justify its cost.

Martinez and Monperrus [23] presented Astor, a library comprising the implementation of three major program repair approaches for the Java platform. The library is also meant to be extended by the research community by adding new repair operators and approaches. The currently supported approaches that originally targeted C programs are:

1. *jGenProg2*: an implementation of *GenProg* for Java [39][18] in which repair operators only consider nearby code, and not the whole codebase as it is the case in *GenProg*.

2. *jKali*: an implementation of the *Kali* approach [33] for Java, which performs repair by exhaustively removing statements, inserting return statements, and switching predicates. Our approach is far from being exhaustive since the predicate switching is highly targeted in terms of location and time.

3. *jMutRepair*: an implementation of the approach presented by Debroy and Wong [10] for the java

platform. *jMutRepair* mutates the relational and logic operators in suspicious *if* condition statements. Since our approach negates predicates at the byte code level (single clause predicates), it practically also mutates relational and logic operators. However, unlike *jMutRepair*, our approach negates the predicates at specific execution instances.

Demarco et al. [11] provided a repair tool, called *Nopol*, which targets buggy *if* conditions and missing pre-conditions in Java programs. *Nopol* uses a test suite that includes a single failing test case; it operates as follows:

1) It considers an *if* statement at location *l* as a repair candidate if forcing it to *true* (or *false*) in all instances would make the failing test case pass. The resulting pair (*l*, *true*) or (*l*, *false*) is used in step 4).
2) It considers a non-*if* statement at location *l* as a repair candidate if skipping it in all instances would make the failing test case pass. The resulting pair would be (*l*, *false*) for skipping the statement or (*l*, *true*) for keeping it.
3) In order to speed up steps 1) and 2), the more suspicious statements are considered first.
4) Given a candidate statement at location *l*, values of the visible variables at *l* are collected for each of its executions and for each test case. The collected program states are encoded as a Satisfiability Modulo Theory (SMT) problem; i.e., an SMT formula is generated which preserves the behavior of the conditional expression for passing tests while modifying it for the failing test case. Lastly, the solution to this SMT (in case it exists) is converted into a source code patch for the faulty program.

*Nopol* differs from our approach on two fronts: 1) the method for determining where and when to repair; and 2) the nature of the provided patch.

An influencing precursor of *Nopol* is *SemFix*, an approach presented by Nguyen et al. [28]. *SemFix* is based on symbolic execution, constraint solving, and program synthesis. Given a candidate repair location *l*, *SemFix* derives constraints on the expression to be injected at *l* in order to make the failing test case pass. Symbolic execution is used to generate the repair constraints, and program synthesis is used to generate the repair patch. Similar to *SemFix*, *DirectFix* [27] and Angelix [26] also aim at synthesizing repairs using symbolic execution and constraint solving; but are more scalable.

Tan and Roychoudhury [38] presented *relifix* an approach for repairing regression bugs. The mutation operators considered are derived by manually inspecting real regressions bugs. The potential repair locations were identified by differencing the current version of the defective program with its previous version, and by considering the *Ochiai* suspiciousness of the locations.

Masri et al. [25] presented an online intrusion and failure detection system that is based on matching against suspicious signatures extracted from execution profiles. The main commonality with the proposed patching approach is that it too can detect when a failure is about to happen, however, it does not provide any means to prevent it.

Pei et al. [30] proposed an approach that exploits contracts such as pre/post-conditions to localize faults and generate repairs in Eiffel programs.

Elkarablieh and Khurshid [14] developed a tool called *Juzi*, within which the user provides a Boolean function that checks whether a given data structure is in a *good* state. The function is invoked at runtime, and in case a corrupt state is detected, the tool performs repair actions via symbolic execution. One of the authors later targeted the repair of the selection conditions in SQL *select* statements [15].

## 7 CONCLUSIONS AND FUTURE WORK

This paper presents an effective patch generation toolset comprising: 1) a highly accurate CBFL technique that identifies failure-causing control dependence chains; and 2) a patch generation mechanism centered on altering the identified control chains. In addition to the provided patch generation capability, the minimal sets of predicates identified by the proposed technique can be valuable for the developers when seeking permanent bug fixes.

As part of future work, we intend to do the following:

1) Improve the accuracy of the classifiers by using program structural information as opposed to state information, or possibly using both types combined.
2) Extend *SinglePredicateSearch* so that it considers combinations of predicates as opposed to single predicates.
3) Explore more sophisticated patterns for negating predicates, possible using a Delta-Debugging-like algorithm or a *genetic* algorithm.
4) Tackle bugs that cannot be fixed via altering control flow; e.g., explore altering data flow chains and devise state altering mutation operators.
5) Devise code synthesis techniques in order to replace the current classifiers with code; i.e., transform the current code patches into actual code repairs.
6) Leverage test case generation in order to alleviate the need for test suites or to improve their quality.

## REFERENCES

[1] Abou-Assi R. and Masri W. Identifying Failure-Correlated Dependence Chains. First International Workshop on Testing and Debugging, TeBug 2011, Berlin, March 2011, pp 607-616
[2] Abreu R., Zoeteweij P., and Van Gemund A. J. C. 2006. An Evaluation of Similarity Coefficients for Software Fault Localization. In PRDC 2006, pages 39–46, 2006.
[3] Agrawal H., Horgan J., London S. and Wong W. Fault localization using execution slices and dataflow sets. IEEE International Symposium on Software Reliability Engineering, ISSRE, pp. 143-151, 1995.
[4] F. Assiri and J. Bieman. Fault localization for automated program repair: effectiveness, performance, repair correctness. The Software Quality Journal. Published Online First, DOI 10.1007/s11219-016-9312-z, March 2016.
[5] George K. Baah, Andy Podgurski, Mary Jean Harrold: Causal inference for statistical fault localization. ISSTA 2010: 73-84.

[6] George K. Baah, Andy Podgurski, Mary Jean Harrold: Mitigating the confounding effects of program dependences for effective fault localization. SIGSOFT FSE 2011: 146-156.

[7] Zhuofu Bai, Gang Shu, Andy Podgurski. NUMFL: Localizing Faults in Numerical Software Using a Value-Based Causal Model. ICST 2015.

[8] Clause J. and Orso A. A Technique for Enabling and Supporting Debugging of Field Failures. Proceedings of the 29th IEEE and ACM SIGSOFT International Conference on Software Engineering, ICSE, pp. 261-270, May 2007.

[9] Dallmeier V., Lindig C., Zeller A. Lightweight Bug Localization with AMPLE. International Symposium on Automated Analysis-Driven Debugging, AADEBUG, pp. 99-103, 2005.

[10] Vidroha Debroy and W. Eric Wong. Using mutation to automatically suggest fixes for faulty programs. In Proceedings of the 2010 Third International Conference on Software Testing, Verification and Validation, ICST '10, pages 65-74, 2010.

[11] F. Demarco, J. Xuan, D. L. Berre, and M. Monperrus, "Automatic repair of buggy if conditions and missing preconditions with SMT," in Proceedings of the 6th International Workshop on Constraints in Software Testing, Verification, and Analysis, CSTVA 2014, 2014, pp.30–39.

[12] Denmat T., Ducassé M. and Ridoux O. Data Mining and Crosschecking of Execution Traces. Int'l Conf. Automated Software Eng., ASE, pp. 396-399, Long Beach, CA, 2005.

[13] Hutchins M., Foster H., Goradia T., and Ostrand T., Experiments of the effectiveness of dataflow- and controlflow-based test adequacy criteria, Proceedings of the 16th international conference on Software engineering, p.191-200, 1994.

[14] B. Elkarablieh and S. Khurshid, "Juzi: A tool for repairing complex data structures," in Proc. 30th Int. Conf. Softw. Eng., 2008, pp. 855–858.

[15] Divya Gopinath, Sarfraz Khurshid, Diptikalyan Saha, Satish Chandra: Data-guided repair of selection statements. ICSE 2014: 243-253.

[16] D. Gopinath, M. Z. Malik, and S. Khurshid. Specification-based program repair using SAT. 17th International Conference on Tools and Algorithms for the Construction and Analysis of Systems (TACAS), pages 173-188, Saarbrucken, Germany, Mar. 2011.

[17] Jones J., Harrold M. J., and Stasko J.. Visualization of Test Information to Assist Fault Localization. In Proceedings of the 24th International Conference on Software Engineering, pp. 467-477, May 2001.

[18] Claire Le Goues, ThanhVu Nguyen, Stephanie Forrest, Westley Weimer: GenProg: A Generic Method for Automated Software Repair. IEEE Trans. Software Engineering 38(1): 54-72 (January/February 2012)

[19] Le Goues C., Holtschulte N., Smith E. K., Brun Y., Devandu P., Forrest S., and Weimer W. The ManyBugs and IntroClass Benchmarks for Automated Repair of C Programs. IEEE Transactions on Software Engineering, vol. 41, no. 12, pp. 1236-1256, 2015.

[20] Liblit B., Aiken A., Zheng A., and Jordan M. Bug Isolation via Remote Program Sampling. Proc. ACM SIGPLAN 2003 Int'l Conf. Programming Language Design and Implementation (PLDI '03), pp. 141-154, 2003.

[21] Liu C., Fei L., Yan X., Han J. and Midkiff S. Statistical Debugging: A Hypothesis Testing-based Approach. IEEE Transaction on Software Engineering, Vol. 32, No. 10, pp. 831-848, Oct., 2006.

[22] Long F. and Rinard M, "An analysis of the search spaces for generate and validate patch generation systems," Proceedings of the 38th International Conference on Software Engineering, pages 702-713, 2016.

[23] Matias Martinez and Martin Monperrus. ASTOR: A Program Repair Library for Java. ISSTA'16, July 18–20, 2016, Saarbrücken, Germany.

[24] Masri, W. Fault Localization Based on Information Flow Coverage. Software Testing, Verification and Reliability (Wiley), Volume 20, Issue 2, pages 121–147, June 2010.

[25] Masri W., Abou Assi R, and El-Ghali M. Generating Profile-Based Signatures for Online Intrusion and Failure Detection. Information and Software Technology (IST) (Elsevier). Vol. 56, Issue 2, Feb. 2014, pages 238-251.

[26] Sergey Mechtaev, Jooyong Yi, and Abhik Roychoudhury. Angelix: Scalable multiline program patch synthesis via symbolic analysis. In Proceedings, ICSE 2016, Austin, TX, May 2016.

[27] Sergey Mechtaev, Jooyong Yi, Abhik Roychoudhury. DirectFix: Looking for Simple Program Repairs. ICSE (1) 2015: 448-458.

[28] Hoang Duong Thien Nguyen, Dawei Qi, Abhik Roychoudhury, Satish Chandra. SemFix: program repair via semantic analysis. ICSE 2013: 772-781.

[29] Chris Parnin, Alessandro Orso: Are automated debugging techniques actually helping programmers? ISSTA 2011: 199-209.

[30] Yu Pei, Carlo A. Furia, Martin Nordio, Yi Wei, Bertrand Meyer, Andreas Zeller. Automated Fixing of Programs with Contracts. IEEE Trans. Software Eng. 40(5): 427-449 (2014).

[31] Pearl J. Causal inference in statistics: An overview. Statistics Surveys Vol. 3 (2009) 96–146, September 2009.

[32] Pearl J. Causality: Models, Reasoning, and Inference. Cambridge University Press, San Francisco, CA, USA, 2000.

[33] Zichao Qi, Fan Long, Sara Achour, and Martin Rinard. An analysis of patch plausibility and correctness for generate-and-validate patch generation systems. In Proceedings of the 2015 International Symposium on Software Testing and Analysis, ISSTA 2015, pages 24-36, New York, NY, USA, 2015.

[34] Renieris M. and Reiss S. Fault localization with nearest-neighbor queries. In Proceedings of the 18th IEEE Conference on Automated Software Engineering, pp. 30-39, 2003.

[35] Rubin D. Estimating Causal Effects of Treatments in Randomized and Nonrandomized Studies. Journal of Educational Psychology, 66:688–701, 1974.

[36] Raul A. Santelices, James A. Jones, Yanbing Yu, Mary Jean Harrold: Lightweight fault-localization using multiple coverage types. ICSE 2009: 56-66.

[37] Gang Shu, Boya Sun, Andy Podgurski, Feng Cao: MFL: Method-Level Fault Localization with Causal Inference. ICST 2013: 124-133.

[38] Shin Hwei Tan, Abhik Roychoudhury. relifix: Automated Repair of Software Regressions. ICSE (1) 2015: 471-482.

[39] Westley Weimer, Thanhvu Nguyen, Claire Le Goues, and Stephanie Forrest. Automatically finding patches using genetic programming. In Proceedings of the 31st International Conference on Software Engineering, ICSE '09, pages 364-374, 2009.

[40] Xie X., Chen T., Kuo F., and Xu B. A theoretical analysis of the risk evaluation formulas for spectrum-based fault localization. ACM Transactions on Software Engineering and Methodology, 2013, Vol. 22, issue 4.

[41] Xie T. and Notkin D. Checking inside the black box: Regression Testing by Comparing Value Spectra. IEEE Transactions on software Engineering, 2005, Vol. 31, issue 10, pp. 869-883.

[42] Xiangyu Zhang, Neelam Gupta, Rajiv Gupta: Locating faults through automated predicate switching. pp 272-281, ICSE 2006.a.

[43] Zhang, X., Gupta, N., and Gupta, R., "Pruning Dynamic Slices with Confidence", Int'l Conf. Programming Language Design and Implementation, PLDI, pp. 169-180, 2006.b.